\documentclass[12pt]{article}
\usepackage{geometry}
\usepackage[superscript,biblabel]{cite}
\usepackage{chngcntr}
\usepackage{setspace}
\usepackage{amsmath}
\usepackage{microtype}
\usepackage{graphicx}
\usepackage{subcaption}
\graphicspath{ {./Figures/} }
 \usepackage{ragged2e}
\begin{document}
\begin{titlepage}
        \vspace*{0.25 cm}
         \huge
         \centering
         \setstretch{1.5}
        \textbf{Machine learning-assisted design of material properties}

         \vspace{1.5cm}
        \LARGE
        \begin{center}
        Sanket Kadulkar,$^{1}$ Zachary M. Sherman,$^{1}$ Venkat Ganesan,$^{1}$ and Thomas M. Truskett$^{1,2}$  
        \normalsize
        \newline
        \par
        
        \textit{$^{1}$ McKetta Department of Chemical Engineering, University of Texas at Austin, Austin, TX, USA} \\
        \par
        \textit{$^{2}$ Department of Physics, University of Texas at Austin, Austin, TX, USA} \\
        \normalsize 
        \centering
        \par
        \end{center}
        \centering
        \large
        E-mail: truskett@che.utexas.edu
          \end{titlepage}

\setlength{\abovedisplayskip}{3pt}
\setlength{\belowdisplayskip}{3pt}

\section*{Abstract}

Designing functional materials requires a deep search through multidimensional spaces for system parameters that yield desirable material properties. For cases where conventional parameter sweeps or trial-and-error sampling are impractical, inverse methods that frame design as a constrained optimization problem present an attractive alternative. However, even efficient algorithms require time- and resource-intensive characterization of material properties many times during optimization, imposing a design bottleneck. Approaches that incorporate machine learning can help address this limitation and accelerate the discovery of materials with targeted properties. Here, we review how to leverage machine learning to reduce dimensionality to effectively explore design space, accelerate property evaluation, and generate unconventional material structures with optimal properties. We also discuss promising future directions, including integration of machine learning into multiple stages of a design algorithm and interpretation of machine learning models to understand how design parameters relate to material properties.

\section{Introduction}

Functional materials are strategically designed to exhibit technologically useful properties. Examples abound, including ionic liquids for carbon capture \cite{Gurkan2010}, nanomaterials for energy storage and catalysis \cite{https://doi.org/10.1002/adma.201706347, SHI201825}, organic materials for photonic applications \cite{C4CS00098F}, and porous materials for hydrogen storage \cite{Chen297}. In most cases, the properties of interest derive from the physical and chemical nature of their constituent building-blocks as well as their spatial organization (i.e., structure).  The characteristics of dopants and additives \cite{Burschka2013, Wang2016} as well as processing conditions affecting structure \cite{Lu2015, https://doi.org/10.1002/smtd.201700229, https://doi.org/10.1002/pssa.201431366, https://doi.org/10.1002/aenm.201700576} impact performance of materials for photovoltaic devices. Microstructure-property relationships have been extensively explored for the design of other material classes including metal alloys \cite{doi:10.1063/1.5021089} and self-assembled block copolymers \cite{Phillip2011, PENDERGAST2013461, Shen2018, Alshammasi2018, Schneider2019, doi:10.1063/1.5128119}. A unifying aspect of materials design is its focus on systematic determination of points in the ``design space'' of experimentally adjustable parameters corresponding to structures and properties optimized for a particular application.

In principle, materials with desirable properties can be discovered using parameter sweeps over the design space. Individual samples must be synthesized or modeled computationally --- and their properties measured or simulated --- for each set of candidate design parameters. Repeating these steps many times with different parameter choices allows one to screen for materials exhibiting targeted properties. However, for most materials of engineering interest, there are many possible parameters to vary, and sweeps covering the corresponding high-dimensional design spaces are impractical. This challenge has been addressed in part by posing materials design as an inverse problem to be solved using methods of numerical optimization to efficiently navigate the design space \cite{Torquato2009, Jain2014, Jaeger2015, Ferguson_2017, Murugan2019, JACKSON2019106, doi:10.1063/1.5145177}. Commonly used algorithms iteratively optimize an objective function formulated based on the desired material properties. At each iteration, the property is measured for the current point in the design space, and the optimizer selects new points to investigate until the algorithm achieves convergence to an optimal solution, within specified tolerances. However, even with sophisticated inverse methods, it may be prohibitively expensive to converge to solutions that satisfy design objectives. 

In this context, machine learning (ML) is beginning to provide powerful new capabilities for the computational design of materials with targeted properties. For example, ML can be used to train a model that replaces the direct computational evaluation of the property of interest, which significantly decreases the time needed for each iteration of an optimization routine \cite{LIU2017159, Ramprasad2017, D0MH01451F, jain_hautier_ong_persson_2016}. There have been several recent reviews that discuss other ways in which ML can be incorporated into an inverse framework to enhance materials design, including using ML to generate new molecules and materials, and to aid the optimizer for prioritized search of design spaces \cite{doi:10.1146/annurev-matsci-082019-105100, polym12010163, doi:10.1021/jacs.0c09105}. Other reviews have focused on ML-assisted design for specific classes of materials, including photonic nanostructures \cite{C9NA00656G, 10.1002/advs.202002923, SoBad2020}, chemical compounds \cite{Schwalbe-Koda2020, Sanchez-Lengeling360, C9ME00039A}, and self-assembled soft materials \cite{Ferguson_2017, JACKSON2019106, doi:10.1063/1.5145177}, as well as how ML might be used for high-throughput experimental investigations \cite{Eyke2021}.

\begin{figure} [tbh!]
\centering
\includegraphics[width = 15.5cm]{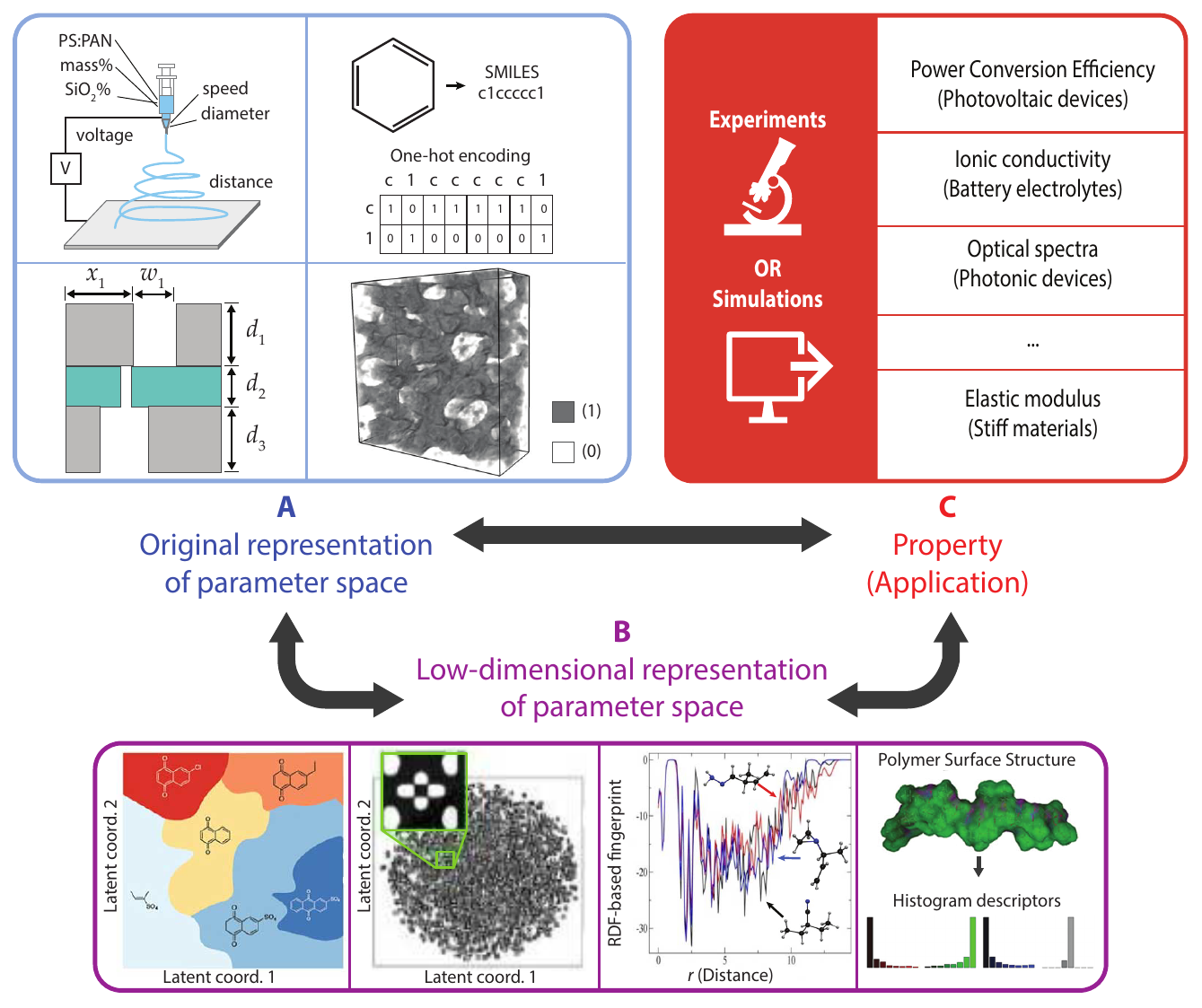}
\caption{\label{fig: Figure1} (a) Detailed, high-dimensional representations of various material systems. (b) Low-dimensional representations of systems derived from their high-dimensional counterparts (a).
(c) Representative material properties that can be measured in experiments or computed in
simulations. Abbreviations: PAN, polyacrylonitrile; PS, polystyrene; RDF, radial distribution function; SMILES, simplified molecular-input line-entry system. Images in panel a (left to
right, top to bottom) adapted with permission from Reference 104, copyright 2020 American
Chemical Society; Reference 46, copyright 2020 The Royal Society of Chemistry; Reference 119,
copyright 2018 American Chemical Society; and Reference 90 (CC BY 4.0). Images in panel b
(left to right) adapted with permission from Reference 53; Reference 51, copyright 2020 AIP
Publishing; Reference 63, copyright 2015 Wiley Periodicals; and Reference 64, copyright 2013
Wiley.}
\end{figure}

Here, we discuss recent advances in ML strategies to design materials with targeted properties. Specifically, we explore how ML approaches vary depending on the representation of the design space, as shown in \textbf{Figure \ref{fig: Figure1}}. Section 2 highlights property design using a low-dimensional representation of the high-dimensional design space. Here, ML is used primarily to reduce the dimensionality of the design space and predict material properties. Section 3 focuses on design solely within the high-dimensional design space, where ML is primarily used to aid an optimizer navigate the space. In Section 4, we outline some promising directions for ML-assisted property design, including combining different ML strategies into a single design framework and improving the interpretability of ML models for design. 

\section{Property design using low-dimensional representations}

To fully characterize a complex material, a high-dimensional representation would be required, including, e.g., positions, orientations, and connectivity of the building blocks. Fortunately, this level of detail is rarely necessary, and material properties can be expressed as functions of far fewer parameters with sufficient accuracy. These parameters form a ``latent space,'' a low-dimensional representation of the design parameters obtained by combining or removing features in the original design space. If the latent space retains the information necessary to compute a material property, then it can serve as a low-dimensional proxy for its high-dimensional counterpart for materials design. This is advantageous because it (i) simplifies the quantitative mapping between the design space and the corresponding property compared to that using the high-dimensional representation and (ii) reduces the number of design parameters an optimizer must modify when navigating the latent space. This section highlights two ways in which ML strategies leveraging low-dimensional latent representations have been implemented to enhance design of material properties. First, we discuss how generative ML models can be used to propose new, nonintuitive material designs with optimal properties directly from the latent space. Second, we explore how ML-based surrogate models quantitatively relate low-dimensional descriptors to the properties.

\subsection{Generative models with latent representation}

\begin{figure} [tbh!]
\centering
\includegraphics[width = 15.5cm]{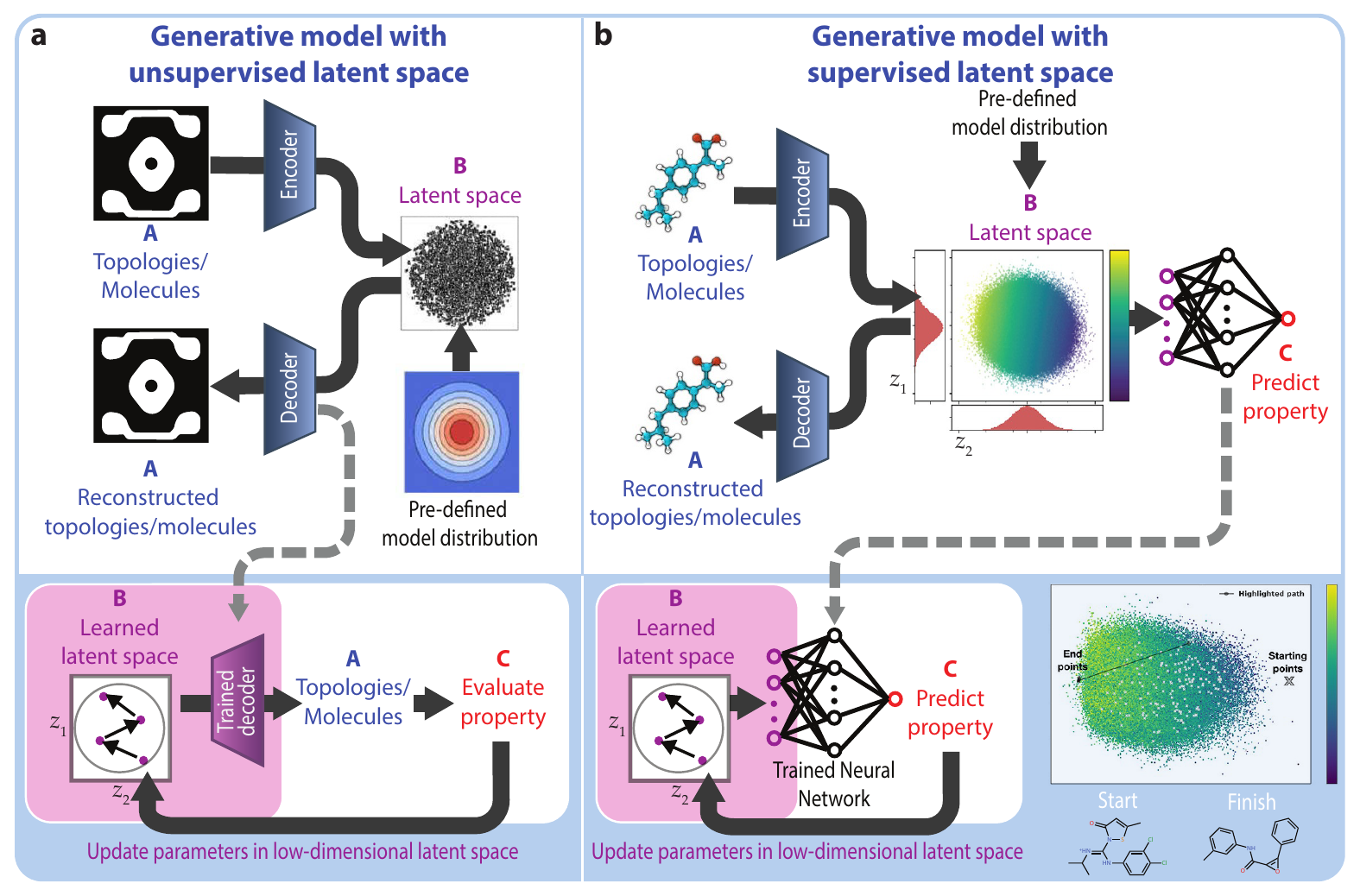}
\caption{\label{fig: Figure2} Machine learning–enabled generation and design of topologies and molecules by latent space sampling. (a) Unsupervised learning of generative latent space representation. A topology or
molecule in the original representation is converted into a vector in the latent space by use of an
encoder. The decoder then reconstructs the corresponding design in the original representation
from the latent representation. Once learning is complete, an iterative method screens the latent
space for target properties, with the trained decoder serving as a generative model. (b) Supervised
learning of latent space representation. The encoder and decoder are trained jointly with a
feedforward neural network–based regressor that predicts a material property from the latent
representation. The trained regressor then predicts material properties directly during iterative
screening of the latent space to design target materials. Panel a adapted with permission from
Reference 50, copyright Walter de Gruyter. Panel b adapted with permission from Reference 53.}
\end{figure}

ML-based models can be constructed for learning a low-dimensional latent space on to which a high-dimensional, detailed design can be projected as well as to reconstruct a design in original representation from any point in the latent space. Inverse schemes can leverage these generative capabilities to search through the latent space, rather than the high-dimensional space, and potentially construct new materials exhibiting desired properties from optimal latent points. Such generative models have been primarily applied for topology optimization and molecular design \cite{doi:10.1021/jacs.0c09105, doi:10.4155/fmc-2018-0358, https://doi.org/10.1002/minf.201700133}. \textbf{Figure \ref{fig: Figure2}} shows two examples of what a high-dimensional representation might look like. In \textbf{Figure \ref{fig: Figure2}\textit{a}}, a multiphase material is characterized by the spatial distribution of its two phases, and the high-dimensional representation consists of a digitized array of pixels, each assigned one of the two phases. \textbf{Figure \ref{fig: Figure2}\textit{b}} shows a molecular structure whose high-dimensional representation contains the positions or connectivity of all atoms, e.g. in the Simplified Molecular Input Line Entry System (SMILES) representation. Generative models project these representations down to just a few latent parameters that retain enough information about the spatial features of the topologies (\textbf{Figure \ref{fig: Figure2}\textit{a}}) or the chemical and structural features of the molecules (\textbf{Figure \ref{fig: Figure2}\textit{b}}). As a result, materials with similar structural motifs typically lie close to each other in the latent space, even if they appear dissimilar or ``far'' from one another in the high-dimensional representation. This feature is useful because we can perform simple operations in the continuous latent space, like perturbations from a single point or interpolations between two points, to propose new high-dimensional representations that may have similar, or perhaps enhanced, properties compared to previously studied materials.

The latent variables are learned by training two separate components (\textbf{Figure \ref{fig: Figure2}}): an encoder which projects a high-dimensional representation of a material to a low-dimensional vector of latent parameters and a decoder or generator which uses a latent vector as input to reconstruct a material in the original high-dimensional representation. The encoder and decoder networks are jointly trained with an unlabeled dataset by minimizing the reconstruction loss, which measures the difference between the original structures in the dataset and the corresponding reconstructed structures. Because the latent representation should facilitate the design of realistic materials, it is helpful if the latent space possesses the property that a random vector fed to the decoder generates physically realistic and meaningful molecules and structures. To ensure this, the learned latent space is also forced to match a predefined target distribution during the training of encoder and decoder. The overall training loss for the model accounts for not only the reconstruction loss but also this latent loss, defined based on the difference between the latent space distribution and the target distribution \cite{10.2307/2236703}.

 There are various generative architectures that have been useful for materials design. With a variational autoencoder (VAE) \cite{kingma2014autoencoding} architecture, the latent space is forced to match a Gaussian distribution. VAEs have been employed for generating material topologies or molecular chemistries for property design in mechanical metamaterials \cite{XUE2020100992, KIM2021109544}, drug discovery \cite{C9SC04026A, C9SC04503A}, and thermoelectric materials \cite{ren2020inverse}. The fixed Gaussian form of the latent space distribution progressively slows the search for optimal solutions as additional constraints on design parameters are introduced, and so strategies which allow for more control of the latent space distribution are desirable for multi-constrained problems. One way to address this challenge is to adopt adversarial autoencoders (AAEs) (a combination of VAE and generative adversarial networks (GANs) \cite{goodfellow2014generative}), an approach that has been sucessfully demonstrated for multi-constrained optimization of the optical response of metastructures within a complex design landscape \cite{Kudyshev2021, doi:10.1063/1.5134792}. \par

In these unsupervised generative strategies, the low-dimensional latent space is discovered independently of any material property of interest. As shown in \textbf{Figure \ref{fig: Figure2}\textit{a}}, the latent space is used to generate structures in the high-dimensional representation, from which a material property can then be characterized directly in experiments or simulations. Though navigating the low-dimensional latent space reduces the number of iterations during an optimization, if measuring the material property is the time-consuming bottleneck, it will still be challenging to converge the optimization. This challenge has been addressed by using supervised methods to train a generative ML model to rapidly compute material properties directly using a point in latent space as input. For example, \textbf{Figure \ref{fig: Figure2}\textit{b}} shows a feedforward regressor trained jointly with an encoder and decoder to learn the latent space representation that best predicts a target material property. The regressor can then be used to quickly compute material properties as an optimizer navigates the latent space. Because this approach completely avoids measuring material properties in simulations or experiments at every iteration, it can significantly accelerate materials design by reducing both the number of iterations and the time per iteration. By training the property predictor jointly with the VAE, the latent variables learned by the model are such that the topological structures or molecular designs exhibiting similar properties will be distributed close together in the latent space. As a result, it is possible to identify principal axes in latent space along which a material property varies, which can greatly simplify the search for optimal materials \cite{WANG2020113377, Yao2021}.  Fully connected neural networks serving as property predictors coupled with a generative VAE model have been successfully employed for design of metamaterials with desired distortion responses \cite{WANG2020113377}, drug-like molecules \cite{C9SC04503A, doi:10.1021/acscentsci.7b00572}, inorganic crystals for thermoelectric materials \cite{ren2020inverse}, metal–organic framework structures for gas seperation applications \cite{Yao2021}, and high thermal conductivity alloys \cite{doi:10.1063/5.0028241}. However, as opposed to the unsupervised training of a VAE architecture, the supervised training of the structure-property regressor component in conjuction with the VAE network requires generation of labeled structural datasets.

\subsection{Forward predictive modeling}

\begin{figure} [tbh!]
\centering
\includegraphics[width = 15.5cm]{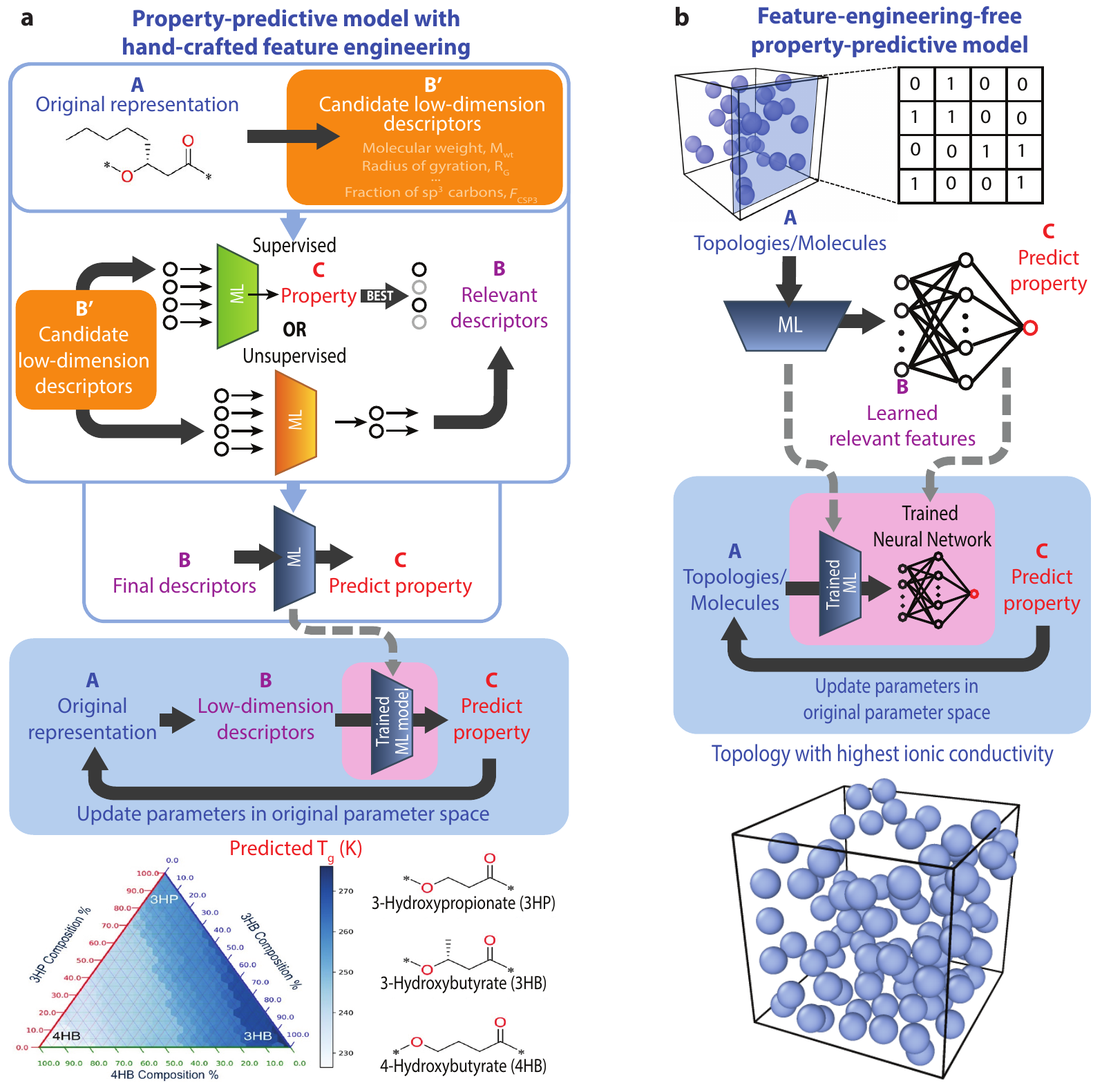}
\caption{\label{fig: Figure3} Two strategies for using ML to predict material properties by leveraging a low-dimensional space of descriptors: (a) First, a pool of candidate descriptors is created by hand, and then ML methods are used to reduce the pool. (b) The low-dimensional set of descriptors is discovered directly during training, without the need to construct a candidate pool. The trained network, obtained
with either strategy, is then integrated into an iterative scheme to design (a) copolymers or (b)
nanoparticle configurations with the desired material properties. Abbreviations: ML, machine
learning; Tg, glass transition temperature; 3HB, 3-hydroxybutyrate; 3HP, 3-hydroxypropionate;
4HB, 4-hydroxybutyrate. Panel a adapted with permission from Reference 68, copyright 2019
American Chemical Society. Panel b adapted with permission from Reference 74, copyright 2021
American Chemical Society.}
\end{figure}

ML has been particularly useful for rapidly predicting material properties. Once a ML model is trained, evaluating a material property using the model is significantly faster than measuring the property in an experiment or computing it in a simulation. This is promising for materials design, as the ML model can replace experiments and simulations to accelerate each iteration of an optimization scheme. To train a ML model, we require a large data set linking inputs to the resulting material properties. However, the choice of input is extremely important. In many cases, using the original, high-dimensional representation of parameter space would require impractically large training sets to adequately sample, and inadequate sampling leads to trained models with inaccurate predictions \cite{https://doi.org/10.1002/gamm.202100003, OO2019, ma14040713}. A more efficient approach is to identify low-dimensional features used as either an input to the ML model or an intermediate layer in the ML architecture. Since the compressed features constituting relevant combinations of the original design parameters help preserve symmetries (e.g. rotational and translational invariance in topologies), this strategy requires much smaller training sets, alleviating the need to explicitly introduce the symmetric variants described in the original representation.

There are two main approaches to finding a low-dimensional representation for material property prediction. The first (\textbf{Figure \ref{fig: Figure3}\textit{a}}) involves first creating a pool of candidate low-dimensional descriptors and then using ML to reduce the pool and find the descriptors most relevant for predicting a target property. In this regard, the hand-crafted features hypothesized to capture most of the material information influencing the property of interest are usually chosen as the candidate descriptors \cite{doi:10.1063/1.5021089, doi:10.1021/acs.jcim.9b00623, https://doi.org/10.1002/polb.24117, C8ME00050F, doi:10.1021/jacs.0c01473, https://doi.org/10.1002/qua.24912, https://doi.org/10.1002/adfm.201301744}. For example, the glass transition temperature $T_g$ of a polymer is a complex property influenced by various structural and compositional features of the polymer. However, instead of a fully-detailed molecular representation, the polymers can be described using physics-inspired descriptors such as molecular weight, radius of gyration, etc (\textbf{Figure \ref{fig: Figure3}\textit{a}}) \cite{doi:10.1021/acs.jcim.9b00807}. The pool can be expanded by using feature-engineering to construct new candidate descriptors through, e.g., arithmetic combinations of the current descriptors in the pool \cite{DAI2020109194, PhysRevMaterials.2.083802, Weng2020, doi:10.1021/acs.jcim.9b00807}. Both supervised and unsupervised methods have been developed to reduce this pool. Supervised learning methods sift through the candidate pool and select only those descriptors that most significantly correlate with the property of interest. Several such methods have proven effective for predicting material properties, including embedded feature selection \cite{LEE20182708}, sure independence screening and sparsifying operator (SISSO) \cite{PhysRevMaterials.2.083802, doi:10.1021/acs.jcim.9b00807}, least absolute shrinkage and selection operator (LASSO) \cite{https://doi.org/10.1111/j.1467-9868.2011.00771.x}, and genetic algorithms for feature selection \cite{Weng2020, ZHANG2020528}. Unsupervised learning methods identify correlations within the descriptor pool and generate a new, smaller set of nonredundant features that are combinations of the original candidate descriptors. Specific unsupervised feature reduction techniques that have been effective for property prediction include principal component analysis (PCA) \cite{Melati2019, Yucel2020, doi:10.1021/acs.jpcb.1c02004}, uniform manifold approximation and projection (UMAP) \cite{Fung2021}, t-distributed stochastic neighbor embedding (t-SNE) \cite{Yuan2020}, and multidimensional scaling \cite{JUNG201917}. It is usually not obvious which of these ML techniques is best for a specific problem, so it can be advantageous to implement several different ML methods and choose the one with the best prediction accuracy \cite{doi:10.1021/acs.jcim.9b00623, ZHANG2020528, doi:10.1021/acs.chemmater.0c02325, doi:10.1021/acsaem.0c02647}. Finally, the trained ML model that links the low-dimensional descriptors to the property of interest can be integrated into an iterative scheme to design materials with optimal properties. As illustrated in Figure \ref{fig: Figure3}A, this strategy has been employed to design random copolymers with targeted values of $T_g$.

The second strategy incorporates discovery of a low-dimensional set of descriptors directly into the training process without requiring an initial pool of hand-crafted features. As illustrated in \textbf{Figure \ref{fig: Figure3}\textit{b}}, the ML model takes the fully-detailed high-dimensional representation as input, and during training, finds the low-dimensional descriptors that best predict the desired material property. This approach requires a supervised learning approach, and the particular set of low-dimensional descriptors that is discovered varies as the property of interest changes. Although these low-dimensional features are abstract and cannot be readily interpreted from a physical standpoint, this strategy is advantageous because the ML model is not constrained to a pool of hand-crafted descriptors which may not capture the information necessary to predict the desired property. Without this limitation, this approach (\textbf{Figure \ref{fig: Figure3}\textit{b}}) can outperform those requiring hand-crafted descriptor pools (\textbf{Figure \ref{fig: Figure3}\textit{a}}) for more accurate property predictions \cite{doi:10.1021/acs.jpcb.1c02004, Kojima2020, goh2017chemception, D0ME00020E, YANG2018278, ROSEN20211578, doi:10.1021/acs.jcim.6b00601, wallach2015atomnet, 8983340}.  Convolutional neural networks (CNNs) and graph convolutional networks (GCNs) are two common architectures for discovering low-dimensional descriptors. CNNs employ convolutional layers to extract a low-dimensional set of spatial features present in a structured data set, like a pixel- or voxel-based digitized image shown in \textbf{Figure \ref{fig: Figure3}\textit{b}}. These features are then linked to the property of interest by means of a fully connected artificial neural network. CNNs have been implemented to accurately predict material properties from the spatial microstructure of nanocomposites \cite{doi:10.1021/acs.jpcb.1c02004, Kojima2020, D0ME00020E}, porous media \cite{KARIMPOULI201989, Wu2019, Wu2019-2}, elastic composites \cite{YANG2018278}, ceramics \cite{KONDO201729}, and molecules \cite{goh2017chemception, wallach2015atomnet}. GCNs, on the other hand, have been used to successfully extract features from the machine-readable molecular graphs representing the arrangement of atoms and bonds in a molecule. They have been employed to predict properties of atomic crystals \cite{PhysRevLett.120.145301}, large organic molecules \cite{doi:10.1021/acs.jcim.0c00687}, and small molecules \cite{doi:10.1021/acs.jcim.6b00601, Kearnes2016, doi:10.1021/acs.jcim.7b00244, ryu2018deeply}. As illustrated in \textbf{Figure \ref{fig: Figure3}\textit{b}}, these reduction strategies can be integrated into an iterative design scheme in the same manner as the approaches of \textbf{Figure \ref{fig: Figure3}\textit{a}}. Such an approach was recently introduced to find microstructures for a nanoparticle-based electrolyte that maximize or minimize ionic conductivity \cite{doi:10.1021/acs.jpcb.1c02004}.

\section{Property design using original representations}

\begin{figure} [tbh!]
\centering
\includegraphics[width = 15.5cm]{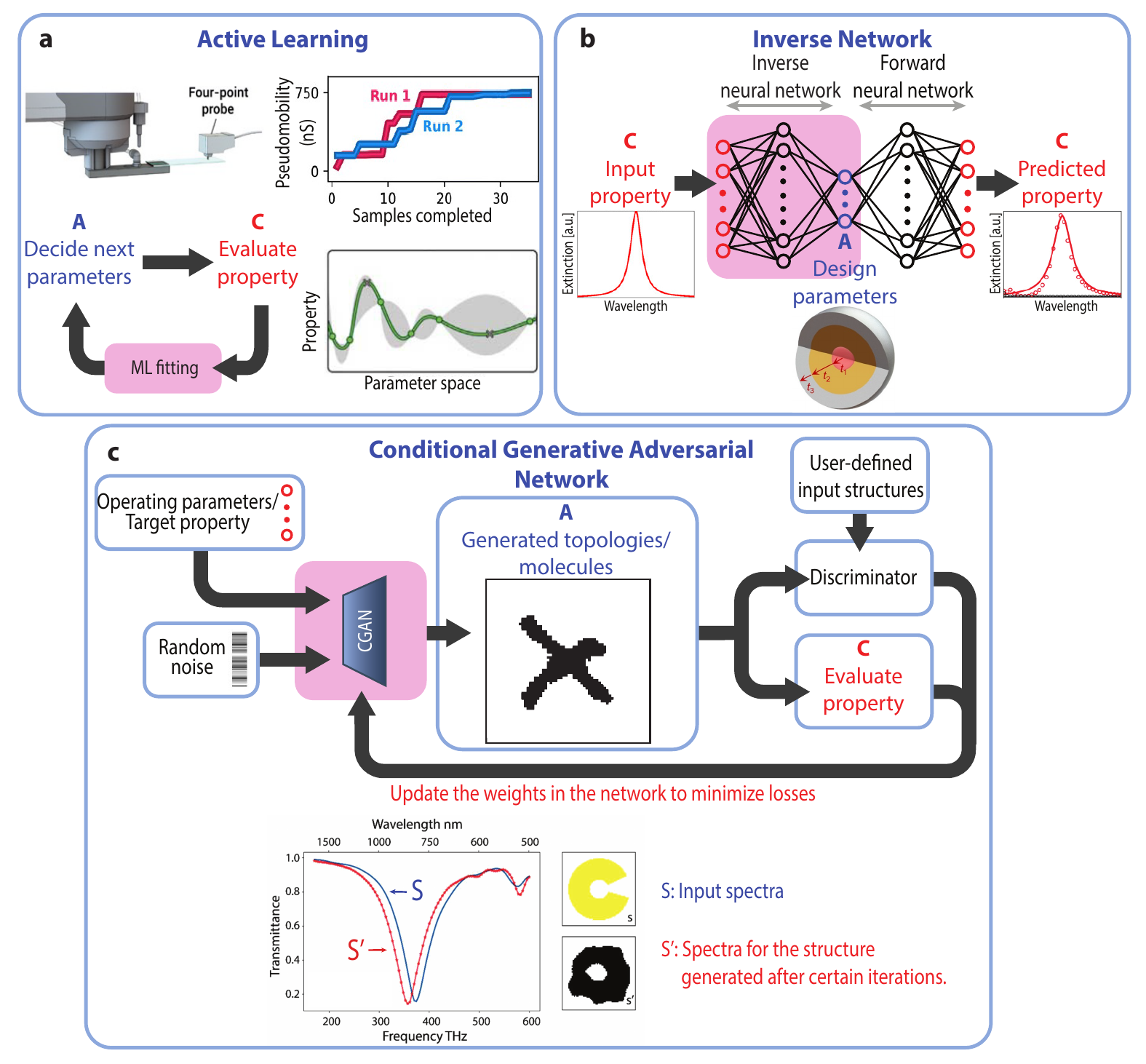}
\caption{\label{fig: Figure4} Machine learning strategies for material design without dimensionality reduction. (a) Starting with a few labeled data points, the property landscape is estimated and used to decide the next set of data points to be evaluated. The predicted property landscape is updated with newly
acquired labeled data points at each iteration, leading to materials with the desired properties.
(b) A tandem neural network with a sequential combination of inverse modeling and conventional
forward modeling networks. The trained inverse network uses a desired property as input and
predicts the corresponding design parameters as output. (c) Given a target property as input,
CGANs are progressively trained to generate structures exhibiting the corresponding target
properties. Starting with the generation of random structural designs during the initial iterations,
the network eventually learns to generate structures optimal for the desired properties.
Abbreviation: CGAN, conditional generative adversarial network. Panel a adapted from Reference
102 (CC BY-NC 4.0). Panel b adapted with permission from Reference 124, copyright 2019
American Chemical Society. Panel c adapted with permission from Reference 131, copyright 2019
American Chemical Society, and from Reference 129, copyright 2018 American Chemical Society.}
\end{figure}

In this section, we highlight ML-assisted design strategies that do not require a compressed, low-dimensional representation of the design space. As a result of the exclusive linkage between the original design parameters and the property of interest, the design of materials using these approaches involves smart navigation of the inherent design space. Although a fixed property-predictive ML model can be trained for accelerated screening even in the absence of a low-dimensional representation, most studies have employed ML to screen candidates and generate designs that achieve desired properties. Here we discuss the three strategies shown in \textbf{Figure \ref{fig: Figure4}}: (a) active learning, (b) inverse neural networks and (c) conditional generative adversarial networks.  

\subsection{Active learning}

Active learning strategies (\textbf{Figure \ref{fig: Figure4}\textit{a}}) are efficient black-box optimization techniques suited for expensive objective functions because they avoid probing uninformative and suboptimal points in the design space. Such strategies are particularly attractive for materials design because they can reduce the total number of times a material property must be evaluated, which is often time-consuming, compared to traditional one-factor-at-a-time approaches. Starting with a small labeled data set, the ML model fits a function to estimate what is known as the ``property landscape'' (i.e., the relationship between the material property of interest and the parameters of the design space). At every iteration, the optimization routine uses this function to suggest a new set of parameters at which to measure the material property, keeping in mind that there is an exploration-exploitation trade-off that must be balanced to avoid restrictively local searches while ensuring efficient optimization. Once the additional property information from the newly selected parameters is known, the ML-estimated property landscape can be refined and the process repeated until convergence of the property values evaluated for the newly sampled design points is achieved. Active learning strategies are typically found to be efficient for exploring low-dimensional parameter spaces (e.g., those less than 20 dimensions) \cite{Frazier2018ATO, Moriconi2020, 10.5555/3013558.3013569}. As a result, the particular active learning strategies employed for material design are typically used to determine the optimal experimental conditions for materials synthesis and processing \cite{doi:10.1021/acsnano.8b04726, MacLeodeaaz8867, DAVE2020100264, doi:10.1021/acsami.0c11667, WEN2019109, MUNSHI2021110119, doi:10.1021/acsaem.0c03201} and to identify the ideal combination of physical parameters to be provided as input in simulations \cite{doi:10.1021/acs.chemmater.9b04830, doi:10.1021/acs.macromol.0c01547} contrary to optimizing high-dimensional design spaces (e.g. structural topologies) to achieve target properties. Feature importance analysis can be performed intermittently to eliminate design parameters that only marginally influence the property, thus reducing the number of dimensions to explore in the subsequent iterations \cite{doi:10.1021/acsnano.8b04726, KHATAMSAZ2021110001}.

 Different techniques for active learning can be categorized based on the choice of ML model used to predict the property landscape and the iterative algorithm employed for determining the next design points to probe. Bayesian optimization is an active learning algorithm widely discussed in previous review articles \cite{doi:10.1146/annurev-matsci-082019-105100, polym12010163, doi:10.1021/jacs.0c09105, Batra2021, Lookman2019} which fits a Gaussian process regression model to the labeled data points at every iteration. In addition to predicting the property landscape, the Gaussian process model also builds an acquisition function based on the predicted mean and variance to guide the location of the next query point. Other approaches \cite{doi:10.1021/acsnano.8b04726, doi:10.1021/acsami.0c11667, WEN2019109, MUNSHI2021110119} have used elastic net regression \cite{https://doi.org/10.1111/j.1467-9868.2005.00503.x} or support vector regression with a radial basis function kernel \cite{Smola2004, JMLR:v11:chang10a, Xue2016, XUE2017532} for predicting the property landscape. For studies employing the ``Design of Experiments" approach, the ML-estimated property landscape can be analyzed by the experiment designer to manually decide the next set of experiments \cite{doi:10.1021/acsnano.8b04726, WEN2019109}. Similarly, evolutionary algorithms such as the differential evolution algorithm \cite{Storn1997} and the metaheuristic cuckoo search algorithm \cite{yang2010cuckoo} have been employed to efficiently explore the property landscape predicted by the ML models \cite{doi:10.1021/acsami.0c11667, MUNSHI2021110119}.

\subsection{Inverse networks}

All of the strategies discussed above navigate through a design space to search for parameters where a material's properties are optimized or closely match those of a target. Inverse networks (\textbf{Figure \ref{fig: Figure4}\textit{b}}) take a different approach and attempt to learn the property-to-design mapping. Where successful, this strategy greatly simplifies materials design since the inverse network can take the target properties as input and immediately output the corresponding design parameters. Inverse networks have been commonly employed for optimizing nanophotonic devices where the physical geometric parameters describing the nanostructure (such as height, length, thickness, etc.) constitute the design parameters, and the resulting optical spectral response of the device is the material property \cite{C9NA00656G, 10.1002/advs.202002923, SoBad2020}. Although the training of an inverse network requires generation of a large data set, it is a one-time cost, and the same network can be repeatedly employed to design materials with different target properties.  \par

The typical architecture for inverse networks is an artificial neural network model trained on material properties as inputs and design parameters as outputs. However, it can be difficult to converge the weights of a stand-alone inverse network during training because the function is multivalued, and many different design points can encode materials with similar properties. To address this issue, a tandem architecture, where a conventional forward-modeling neural network is appended to the inverse network, was introduced (\textbf{Figure \ref{fig: Figure4}\textit{b}}) \cite{doi:10.1021/acsphotonics.7b01377}. This tandem neural network architecture could be successfully trained and offered excellent prediction accuracy when designing photonic structures for a target electromagnetic response. The forward network was first trained independently and remained frozen during the training of the tandem architecture. The weights in the inverse network were trained by minimizing the error between the real property input to the tandem network and the output property predicted by the tandem network. Other studies \cite{Malkiel2018, PhysRevB.101.195104, 8864295, doi:10.1063/5.0055733} have similarly reported that this strategy helps training convergence, despite the inverse network itself being multivalued, because the training losses are defined only by the property loss and not on the error between the predicted and actual design parameters. The tandem inverse architecture was also found to be effective for simultaneously predicting a combination of discrete design parameters (materials indexed by numbering) and continuous structural parameters (thicknesses) displaying a targeted optical spectrum \cite{doi:10.1021/acsami.9b05857, QIU2021126641}. 

The limitations of the tandem architecture to handle the non-unique response-to-design mapping for systems with low-dimensional design parameters were also discussed in a recent study \cite{cryst10020125}. Another strategy to resolve this nonuniqueness involves a stand-alone inverse network with design parameters modeled as multimodal distributions rather than discrete values \cite{Unni2020}. The output from the inverse network now represents weighted multiple design solutions for the input material property; however, the approximate number of degenerate solutions needs to be known in advance. To date, inverse networks have been primarily applied to design nanostructured photonic systems. Their applicability for designing other classes of materials, though promising, remains largely unexplored.

\subsection{Conditional Generative Adversarial Networks}

For material systems with high-dimensional parameter spaces (e.g., nanostructured topology design or molecular design), identifying a low-dimensional latent space is a potential avenue to speed up the optimization. However, the search through either the latent space (for generative models as shown in \textbf{Figure \ref{fig: Figure2}\textit{a}}) or the original design space (for property-predictive models as shown in \textbf{Figure \ref{fig: Figure2}\textit{b}}) will be driven by a separate optimization algorithm. In such cases, the sheer number of degrees of freedom can hinder the discovery of optimal designs. This challenge can be mitigated by utilizing conditional generative adversarial networks (CGANs) as the generative model, trained to bias the generation towards optimal structures with desirable properties. Conditional generative adversarial networks (CGANs) have been used mainly for inverse problems in the design of molecular species \cite{PPR:PPR95844} and structural topologies \cite{doi:10.1021/acs.nanolett.8b03171, SoRho+2019+1255+1261, doi:10.1021/acsnano.9b02371, Kimeaax9324}. The distinguishing feature of CGANs (\textbf{Figure \ref{fig: Figure4}\textit{c}}) is that they combine training of the networks and optimization of design parameters in a single step. Compare this to other ML-based generative models, such as VAEs shown in \textbf{Figure \ref{fig: Figure2}}, which separately train ML models and then use them in an iterative optimization scheme.  In a CGAN, the weights in the network are updated at each iteration to both improve generative capabilities as well as progressively shift the generated structures toward those exhibiting target properties. In this way, the CGAN avoids generating stochastic structures with suboptimal properties, focusing only on reliably generating structures with properties similar to the user-defined target. This combined process of updating network weights to generate structures and then computing the material property is performed repeatedly until convergence to a specified tolerance. CGANs do not require a large labeled data set beforehand, but material properties at each iteration have to be evaluated from their original representation, either explicitly using experiments or simulations or leveraging a separate ML model for property prediction.

 A typical CGAN architecture (\textbf{Figure \ref{fig: Figure4}\textit{c}}) consists of two components: a generator network which creates structures distributed over the design space and a discriminator network which distinguishes the generated designs from the user-defined (real) designs. During each training step, the weights in the generator are updated based on two different losses. First, losses based on the distance between the evaluated properties of the generated designs and the target properties ensures biasing of the generator toward designs with desired properties over several iterations. Second, losses quantified by the discriminator based on the difference between the distributions of the generated designs and the fixed distribution of designs in the user-defined data aim to train the generator to produce a wide distribution of realistic designs, avoiding local optima in the design space \cite{doi:10.1021/acs.nanolett.8b03171, doi:10.1021/acs.nanolett.9b01857}. Reinforcement learning, in conjunction with the discriminator, can be used as an alternative strategy to the CGAN architecture in \textbf{Figure \ref{fig: Figure4}\textit{c}} to bias generated structures to those with desired properties \cite{PPR:PPR95844}. 
 
 In many cases, we would like to maximize (or minimize) a material property, which makes it difficult to evaluate property losses using standard loss functions that compare two properties. One solution is to, at each iteration, define the target property for the loss as the highest (or lowest) value among all of the previously sampled designs \cite{doi:10.1021/acs.nanolett.9b01857}. The targeted design of certain material properties using CGANs also requires as input to the generator the conditional vector, which comprises key operating parameters. For example, studies focused on design of high-efficiency optical nanostructures at different wavelengths and deflection angles have reported CGANs with the corresponding wavelength and angle pair as inputs to the generator \cite{doi:10.1021/acsnano.9b02371, doi:10.1021/acs.nanolett.9b01857}.

\section{Future Directions}

The strategies discussed in this review highlight how ML offers some efficient solutions for addressing key challenges in inverse approaches to materials design. Some of the methods have been developed only very recently and have great potential for future use in different stages of the design workflow. We specifically discuss opportunities to combine two or more ML strategies in a single inverse workflow (Section 4.1) as well as strategies that interpret ``black-box" predictions of ML models to provide fundamental insight on the relation between design parameters and material properties (Section 4.2).

\subsection{Combining strategies}

Most of the inverse strategies discussed here employ ML methods to assist with a single phase of the design scheme. However, it may be advantageous to combine multiple ML techniques, each enhancing a different part of the design process. This could significantly accelerate design of materials, but many details, such as which ML strategies are compatible with one another as well as the application-specific training requirements are not presently known.

One strategy is to train a ML model to predict the properties corresponding to materials encoded in the latent space, e.g., discovered from unsupervised learning of generative models such as VAEs or GANs. Optimization can proceed quickly in the low-dimensional latent space, and each iteration is fast using the ML model to evaluate properties, an improvement over the scheme in \textbf{Figure \ref{fig: Figure2}\textit{a}}, which requires explicit simulations or experiments. Pretrained convolutional neural networks \cite{Kudyshev2021, doi:10.1063/1.5134792} and Gaussian process regression models \cite{KIM2021109544} quantitatively linking the designs in the original representations to the property of interest can accelerate property evaluation during the iterative search of the unsupervised latent space. However, in some cases it might not be practical to generate the required training sets. In these cases, active learning strategies, suitable for low-dimensional design spaces, can be used to query the compressed latent variables for identifying designs that score highly based on the desired material properties \cite{XUE2020100992, C9SC04026A}. Such a strategy circumvents generating large labeled training data beforehand.   \par 

A particularly interesting approach involves combining autoencoder networks with either a feedforward, property-predictive model (for the forward problem) or an inverse network (for the inverse problem) to reduce the computational expense associated with the design. In this strategy, recently applied to designing optical metasurfaces \cite{Kiarashinejad2020, Ma_2020}, the strong correlations present within the structural features as well as the optical response features are exploited to reduce the dimensionality of both the design and property space using autoencoder networks. This one-to-one mapping between the design parameters and property in their reduced spaces is beneficial for design of materials as it allows for employing inverse networks without nonuniqueness, and further alleviates the network-size issues for both the forward and inverse networks.

\subsection{Interpretability of models}

Although ML-based models facilitate discovery of materials with desired properties, the learned structure-property relations are often difficult to interpret. However, it is possible to develop methods that examine trained ML models to elucidate new correlations between the design parameters and the properties of interest. This could provide valuable physical insights that facilitate the experimental realization of material designs within realistic constraints. \par

Several techniques have been established to interpret the ML models used for material design, especially for ML models used for forward property prediction. For example, various feature importance scores that quantify the significance of individual descriptors on a material property can be computed using Shapley additive explanations (SHAP) \cite{MAULANAKUSDHANY2021190, Chen2021}, Gini important analysis \cite{doi:10.1021/acs.jcim.9b00623}, and mean decrease accuracy (MDA) \cite{doi:10.1021/acs.macromol.0c00104}. For deep artificial neural networks trained to predict material properties, the design parameters that strongly correlate with the material property can be identified by analyzing the weights of the trained networks \cite{SABANDO2019105777, LEE2021109260}. Unsupervised data-driven approaches such as principal component analysis can also quantify correlations between hand-crafted features and material properties \cite{doi:10.1021/acs.jpcb.1c02004,doi:10.1063/1.5049849,doi:10.1063/1.5049850}. \par

Although the predictive performance of end-to-end forward predictive models (\textbf{Figure \ref{fig: Figure3}\textit{b}}) with no hand-crafted features exceeds those using hand-crafted features, interpreting such ML techniques is more difficult. To that end, saliency mapping is a visualization technique that can be leveraged for interpretation of trained CNN models \cite{simonyan2014deep}. Saliency maps highlight the regions in the digitized image of a structure correlating with the corresponding structure-dependent property based on the learning of the trained CNN model. These techniques pertaining to interpretation of trained CNN models have been applied to identify the underlying microstructural features influencing the corresponding macroscopic properties of materials such as ionic conductivity in ceramics \cite{KONDO201729} and photovoltaic performance in thin-film organic semiconductors \cite{Pokuri2019}. Similarly, the integrated gradients method can interpret trained graph neural networks (e.g. for molecular design) by quantifying the strength of the contributions of the atom and atom-pair features towards the material property \cite{sundararajan2017axiomatic, Dai2021}.

Besides interpreting trained ML models to discover underlying physical laws governing a material property, machine learning techniques can also be used to train accurate yet simple predictive models that are easy to interpret. For example, a recent study \cite{Desai2021} reported training neural networks with adjustable parameters quantifying the complexity of the learned functions to find accurate and physically interpretable expressions for predicting a material property of interest. Similarly, a highly interpretable linear ML model for predicting material properties called factorized asymptotic Bayesian inference hierarchical mixture of experts was also reported \cite{iwasaki2019materials}. The prediction accuracies of this model were comparable to difficult-to-interpret nonlinear models, such as neural networks or support vector machines.

\section{Conclusions}

Machine learning has recently emerged as an effective tool for making materials design problems tractable from a time and resource standpoint. In this review, we have discussed different ML-assisted strategies implemented for inverse design of material properties. Broadly, these strategies employ ML models to either directly or indirectly assist with the accelerated identification of optimal design points potentially yielding the target properties. \par 

For certain design problems, the main information in the original high-dimensional design spaces can be effectively captured with a compressed, low-dimensional representation. In this regard, ML-inspired generative models serve as a means to generate new molecular to topological designs from the compressed latent vectors to identify materials with desired properties. Also, simplified training of the property-predictive ML models with the low-dimensional data allows for accelerated screening of the design space. For systems without the existence of a low-dimensional representation, in addition to the property-predictive modeling, the ML-guided design strategies focus on employing ML methods explicitly to search the design space efficiently. These methods include active learning strategies to sequentially explore new design points based on a surrogate property landscape continually updated as the additional information flows in, backward mapping from target property to design parameters using inverse networks, and generative models trained to bias the generation of designs towards those exhibiting desired properties. \par

The progress reviewed here highlights the applicability of ML techniques for designing materials with tailored properties. Promising future directions, including combining ML strategies for new integrated design approaches and developing improved methods for interpreting trained ML models, further underline the role that ML will continue to play in addressing challenges posed by this rich and important class of inverse problems.

\section*{DISCLOSURE STATEMENT}
The authors are not aware of any affiliations, memberships, funding, or financial holdings that might be perceived as affecting the objectivity of this review. 

\section*{ACKNOWLEDGMENTS}
This work was primarily supported by the National Science Foundation through the Center for Dynamics and Control of Materials: an NSF MRSEC under Cooperative Agreement No. DMR-1720595. The authors acknowledge an Arnold O. Beckman Postdoctoral Fellowship (ZMS) and the Welch Foundation (Grant Nos. F-1599 and F-1696) for support.

\par

\end{document}